\documentclass[twocolumn,showpacs,preprintnumbers,amsmath,amssymb,superscriptaddress,floatfix]{revtex4}

\usepackage{graphicx}
\usepackage{dcolumn}
\usepackage{bm}

\begin{document}

\preprint{APS/123-QED}

\title{Spatial structure of Mn-Mn acceptor pairs in GaAs}

\author{A.~M.~Yakunin}
 \affiliation{COBRA Inter-University Research Institute, Eindhoven University of Technology, P.O.Box 513, NL-5600MB Eindhoven, The Netherlands}
 \author{A.~Yu.~Silov}
 \affiliation{COBRA Inter-University Research Institute, Eindhoven University of Technology, P.O.Box 513, NL-5600MB Eindhoven, The Netherlands}
 \author{P.~M.~Koenraad}
 \affiliation{COBRA Inter-University Research Institute, Eindhoven University of Technology, P.O.Box 513, NL-5600MB Eindhoven, The Netherlands}
\author{J.-M.~Tang}
\affiliation{Optical Science and Technology Center and Department of Physics and Astronomy,
University of Iowa, Iowa City, IA 52242, USA}
\author{M.~E.~Flatt\'{e}}
\affiliation{Optical Science and Technology Center and Department of Physics and Astronomy,
University of Iowa, Iowa City, IA 52242, USA}
\author{W.~Van~Roy}
\affiliation{IMEC, Kapeldreef 75, B-3001 Leuven, Belgium}
\author{J.~De~Boeck}
\affiliation{IMEC, Kapeldreef 75, B-3001 Leuven, Belgium}
\author{J.~H.~Wolter}
\affiliation{COBRA Inter-University Research Institute, Eindhoven University of Technology, P.O.Box 513, NL-5600MB Eindhoven, The Netherlands}

\date{\today}

\begin{abstract}
The local density of states of Mn-Mn pairs in GaAs is mapped with cross-sectional scanning tunneling microscopy and compared with theoretical calculations based on envelope-function and tight-binding models. These measurements and calculations show that the crosslike
 shape of the Mn-acceptor wave-function in GaAs persists even at very short Mn-Mn spatial separations. The resilience of the Mn-acceptor wave-function to high doping levels suggests that ferromagnetism in \hbox{GaMnAs} is strongly influenced by impurity-band formation.  The envelope-function and tight-binding models predict similarly anisotropic overlaps of the Mn wave-functions for Mn-Mn pairs.  This anisotropy implies differing Curie temperatures for  Mn $\delta$-doped layers grown on differently oriented substrates.
\end{abstract}
\pacs{71.55.Eq, 73.20.-r, 75.50.Pp, 75.30.Hx}
\maketitle
The properties of dilute ferromagnetic semiconductors, such as Ga$_{\rm 1-x}$Mn$_{\rm x}$As, depend sensitively on the nature of the spin-polarized holes introduced into the host by the magnetic dopants\cite{SpintronicsBook,Nitin-review}.  Considerable controversy persists about the nature of isolated magnetic dopants in many semiconducting hosts (e.g. Mn dopants in GaN). Measurements of 
the local density of states (LDOS) near an individual Mn substituted for a Ga atom  in GaAs (Mn$_{\rm Ga}$) by cross-sectional scanning tunneling microscopy (X-STM)\cite{Manipulation} have resolved this question for Ga$_{\rm 1-x}$Mn$_{\rm x}$As: there is a hole state bound to the Mn dopant, yielding a Mn$^{2+}3d^{5}+\text{hole}$~complex\cite{PL_Mn_1964,Schneider_Mn+h,Averkiev_Mn_J1-4,Linnarsson_Mn_ExSp} that produces an  extended, highly anisotropic LDOS. The anisotropic shape of the bound hole state at distances $\gtrsim 1$~nm, originating from the cubic symmetry of GaAs\cite{YakuninPRL_Mn}, suggests highly anisotropic Mn-Mn interactions\cite{Flatte_Mn,Guo_2004_Mn-Mn,Fiete_Mn-Mn}.
 
The ferromagnetic properties of Ga$_{\rm 1-x}$Mn$_{\rm x}$As, however, depend on whether this shape persists for concentrations $x$ of Mn impurities for which Ga$_{1-x}$Mn$_x$As is ferromagnetic ($x\gtrsim 0.01$)\cite{Dietl_Mn_Science}. Popular models of Ga$_{1-x}$Mn$_x$As assume that holes residing in a bulk GaAs-like valence band, and thus evenly distributed throughout the material, mediate the ferromagnetic interaction among Mn spins\cite{Dietl_Mn_Science}. However, angle resolved photoemission spectroscopy\cite{Okabayashi_AnglRes_PhEm} observes an impurity band near~$E_F$ and infrared absorption measurements reveal a strong resonance near the energy of the Mn acceptor level as well as deeper in the band-gap of GaAs\cite{Nagai_SpinPolAdsorb,Singley_Infrared}. Furthermore, recent Raman scattering experiments have shown that a Mn$^{2+}3d^4$ configuration partially occurs  for $x>0.02$\cite{SapegaMn}. 
Recent theories suggest significant modifications in the ferromagnetic properties of Ga$_{1-x}$Mn$_x$As if the holes reside in a strongly disordered impurity band\cite{Bhatt-PRL-2001,Timm-PRL-2002,Fiete_Mn-Mn,Bouzerar_Mn,Fiete_GaMnAs-cm}. If the Mn density is near the metal-insulator transition, and individual Mn dopants states are weakly hybridizing with each other, then the inhomogeneous hole density of the impurity band measured by X-STM near an individual Mn dopant should closely resemble that of an isolated neutral Mn, only weakly perturbed by Mn-Mn interactions.

\begin{figure}
  \includegraphics{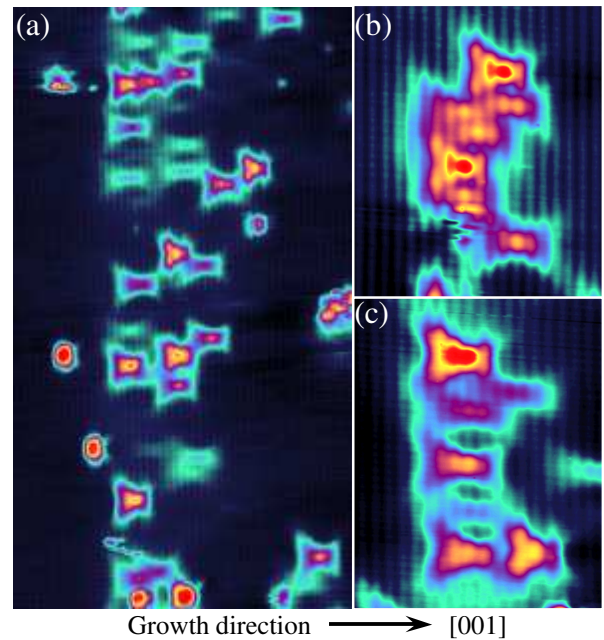}
  \caption{(Color online) Room-temperature X-STM constant-current image of a section of a Mn $\delta$-doped layer with intentional Mn concentrations of  $3\times10^{13}$~cm$^{-2}$: (a)~$51\times29$~nm$^2$; (b)~$12\times10$~nm$^2$; (c)~$13\times10$~nm$^2$. Images were acquired at a sample bias of $U_s=+1.5~V$ on a cleaved (110) surface.}\label{delta_layers}
\end{figure}

\begin{figure}
  \includegraphics{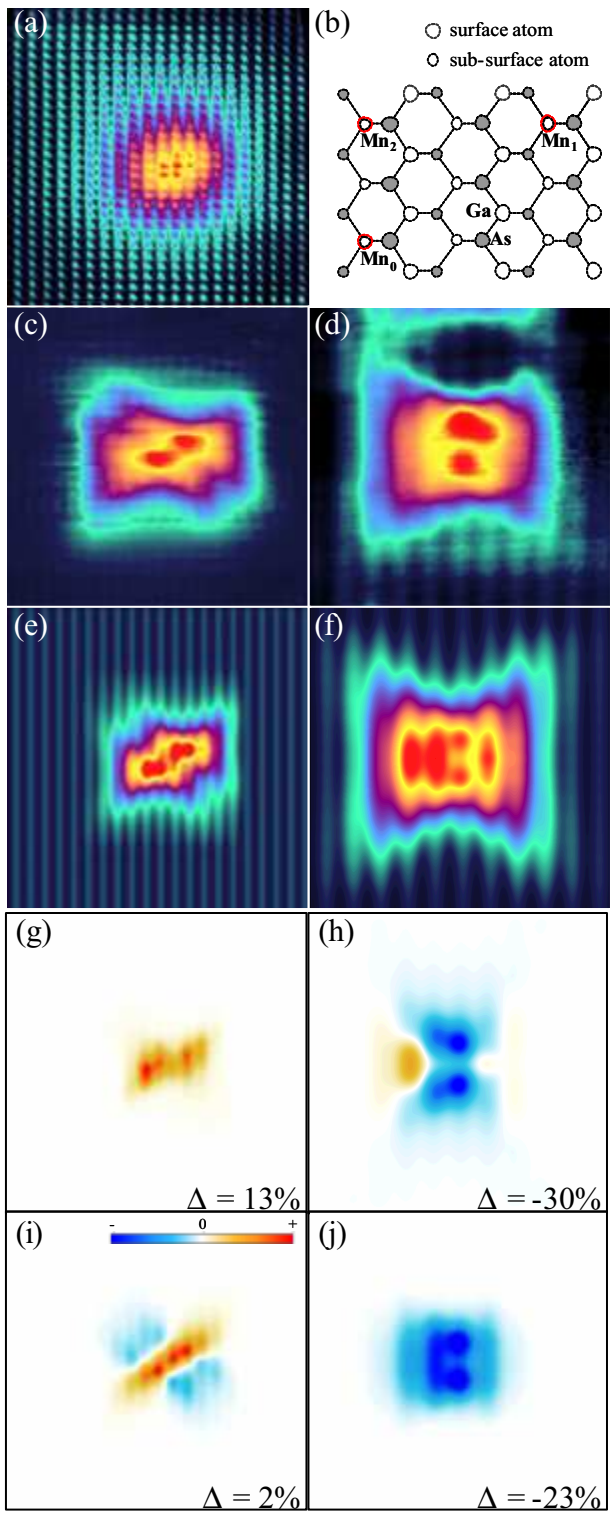}
  \caption{(Color online) Topography of Mn-Mn pairs: (a)~$12\times12$~nm$^2$, X-STM image of two ionized Mn separated by $1.4$~nm  ($U_s=-0.6~V$); (b)~ Schematic of the (110) surface showing the location of subsurface Mn; (c)~same area as (b), imaged with $U_s=+1.1~V$ so the Mn are neutral; (d)~$6\times6$~nm$^2$, X-STM image of two Mn separated by $0.8$~nm ($U_s=+1.55~V$); (e,f)~Calculation with TBM of (c,d), with parallel spins; (g,h)~Difference between Mn-Mn pairs with parallel spins and non-interacting single Mn's; (i,j)~Same as (g,h) but antiparallel spins.}\label{2Mn}
\end{figure}

Here we present experimental evidence that the shape of a Mn acceptor in a Mn-Mn pair remains anisotropic and retains the crosslike shape of a single Mn even when the dopants are separated by only 0.8~nm, which is the typical distance in Ga$_{\rm 0.96}$Mn$_{\rm 0.04}$As. 
The overlapping Mn wave-functions at such short Mn-Mn separations are exceptionally hard to disentangle  in bulk \hbox{GaMnAs}\cite{Grandidier-XSTM-APL,Ohno-XSTM-APL,NRL-XSTM-PRB}. 
In the Mn $\delta$-doped layers of Fig.~\ref{delta_layers}, however, isolated pairs and dense groups of Mn acceptors at these distances can be analyzed separately from surrounding dopants.

The measurements were performed on several samples using chemically etched
tungsten tips. The $\delta$-doped layers were grown at 370~$^{\text{o}}$C by molecular beam epitaxy on a 100~nm GaAs buffer on a Zn-doped (001) GaAs substrate. The high growth temperature was chosen to suppress the appearance of structural defects such as As antisites, which would complicate Mn identification. The higher growth temperature also led to increased segregation, which broadened the $\delta$-doped layers of Fig.~\ref{delta_layers}. Despite the high growth temperature a low density ($<10^{17}$~cm$^{-3}$) of As antisites was observed and clearly identified as charged $n$-type donors (not shown in the figure). 
The $\delta$-doped layers themselves clearly showed $p$-type conductivity in tunneling $I(V)$ spectroscopy. The topographies were measured with a room temperature UHV-STM~($P<2\times10^{-11}$~torr) on an \emph{in situ} cleaved (110) surface.

Figure~\ref{2Mn}(a) shows the electronic topography of one of the pairs in the ionized state (the other observed pairs are similar). In the ionized configuration, it is impossible to distinguish locations of the two dopants. The potential from the double charge of the two ionized dopants induces an apparent round elevation 1.7 times larger than that of a single ionized Mn under the same imaging conditions\cite{Ohno-XSTM,YakuninPRL_Mn}. In the neutral configuration, however the presence of two dopants can be clearly identified. Further two examples of close, clearly identifiable Mn-Mn pairs illustrate the resilience of the Mn wave-function to interaction with nearby Mn dopants [Fig.~\ref{2Mn}(c,d)]. 

A schematic model of the GaAs (110) surface is shown in Fig.~\ref{2Mn}(b). The surface locations of the two Mn of the first pair [shown in Fig.~\ref{2Mn}(a,c,e,g,i) and separated by $1.4$~nm] are indicated by Mn$_{\rm 0}$ and Mn$_{\rm 1}$; the Mn atoms themselves are located in the fifth sub-surface layer, and are well separated from neighboring Mn dopants. The other pair [shown in Fig.~\ref{2Mn}(d,f,h,j) and separated by $0.8$~nm] is indicated in figure~\ref{2Mn}(b) by Mn$_{\rm 0}$ and Mn$_{\rm 2}$, and is likewise in the fifth sub-surface layer as well. This combination has the smallest separation of those Mn-Mn pairs that we were able to identify. These measurements show that in the neutral state, the wave-functions of the two Mn acceptors retain their crosslike shape even when they are separated by a distance smaller than the wave-function's effective Bohr radius $a_0\approx 0.9$~nm.

Figure~\ref{2Mn}(e,f) shows the topography calculated with the tight-binding model (TBM)\cite{Flatte_Mn,YakuninPRL_Mn} for the two pairs shown in Fig.~\ref{2Mn}(c,d), and for Mn spins parallel to each other. The calculation is averaged over the orientation of the two parallel Mn spins relative to the crystal's lattice. Qualitatively the crosslike shape is clearly evident, and the agreement between the calculations and the measurements is as good as found for a single Mn dopant in Ref.~\onlinecite{YakuninPRL_Mn}. The TBM is based on the deep level model of
Vogl and Baranowski\cite{Vogl} and is applied to a bulk-like Mn acceptor. The dangling $sp^{3}$-bonds from the nearest-neighbor As hybridize with the Mn $d$-states of $\Gamma_{15}$ character. The antibonding combination of these becomes the Mn acceptor state. Coupling to the $d$-states of $\Gamma_{12}$ character is weak, and hence neglected. The hybridization strength is fully determined by the acceptor level
energy. 
 
Comparison with theoretical calculations based on the TBM permits a quantitative evaluation of the effect of the Mn-Mn interaction on the measured topography. The topography for a ``non-interacting baseline'' is constructed by adding together the topography of two single, isolated Mn displaced by the pair separation. This baseline is then subtracted from calculations of Mn pairs with differing spin orientations. Figure~\ref{2Mn}(g,h) shows the difference between Fig.~\ref{2Mn}(e,f) and the non-interacting baseline. The quantity $\Delta$ is the ratio of the largest difference shown in Fig.~\ref{2Mn}(g,h) to the largest value of the topography for the non-interacting baseline, and even for the close pair is less than $1/3$.  In previous work the spectral and spatial differences between dopant spin pairs with parallel and antiparallel spins were predicted\cite{Flatte_Mn,Flatte-Reynolds}. Here the spectra could not be measured with sufficient resolution to distinguish the pair spin orientations, and the differences expected between parallel and antiparallel spins are small. Figure~\ref{2Mn}(i,j) shows the same as Fig.~\ref{2Mn}(g,h), except that the Mn spins are antiparallel. The differences between parallel and antiparallel are of the order of 10\%, which is not resolvable in our measurements.

Now that the robustness of the anisotropic crosslike shape of the Mn hole wave-function has been clearly established, we explore the implications for spin-spin coupling mediated by these hole wave-functions in an impurity band.  The experimental data acquired with STM is a two-dimensional slice along a (110) plane of the entire three-dimensional wave function.  As a result, any estimation of the directional dependance of the wave-function overlap taken directly from the STM experiment would be incorrect. Instead we  quantify the directionally dependent overlap of the wave-functions by calculating the bulk-like Mn-acceptor wave-function within a four-band Luttinger-Kohn envelope-function model (EFM) as well as the tight-binding model. The EFM uses the zero-range potential model\cite{ZeroRange}, including a cubic correction as suggested in Ref.~\onlinecite{YakuninPRL_Mn}. The ground state of the Mn acceptor can be approximated as four-fold degenerate with a total momentum of the valence hole $F=3/2$ and has the symmetry of the top of the valence band $\Gamma_{8}$\cite{BirPikus}. We neglect possible effects caused by the presence of the (110) surface and quantum spin effects from the exchange interaction between the Mn$^{2+}3d^{5}$ core and the hole.

\begin{figure}
  \includegraphics[width=7.5cm]{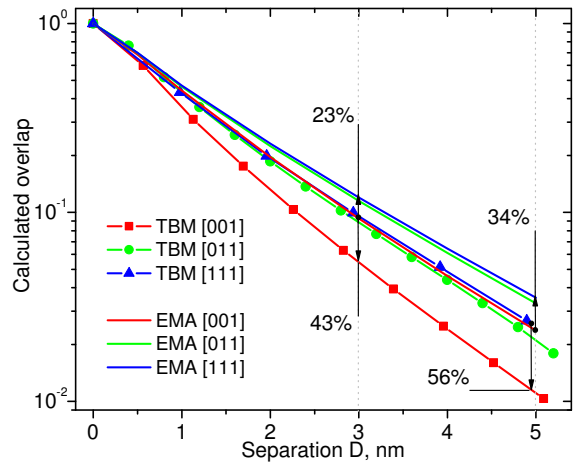}
  \caption{(Color online) Separation-dependent overlap of non-interacting Mn acceptors calculated for separations along three crystallographic directions using the envelope-function model (EFM) and the tight-binding model (TBM).}\label{2Mn_directions}
\end{figure}

\begin{figure}
  \includegraphics[width=8.2cm]{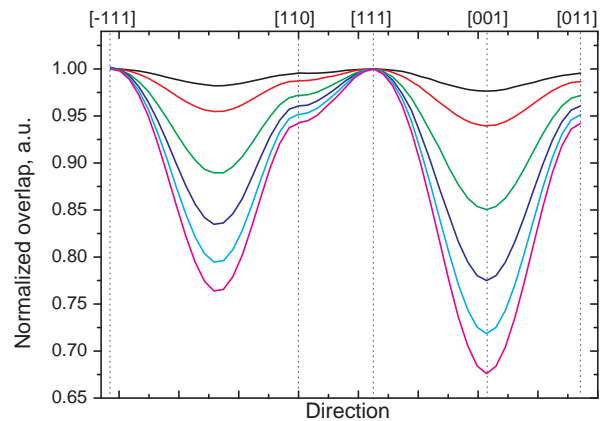}
  \caption{(Color online) Directional dependance of the Mn wave-function overlap for Mn-Mn separations \emph{D} calculated in the  EFM. The curves are normalized to the maximum value \emph{N} given in Table~\ref{tab:table}.} \label{2Mn angle}
\end{figure}

The calculated radial dependance of the overlap of non-interacting Mn wave-functions for three crystallographic directions is presented in Fig.~\ref{2Mn_directions}. The graph shows a nearly exponential decay of the overlap integral with separation, but characterized by a directionally-dependent decay constant. Thus the anisotropy of the overlaps increases at larger separations between the Mn dopants. The calculated directional dependence of the overlaps of non-interacting Mn wave-functions for various Mn-Mn separations is shown in Fig.~\ref{2Mn angle}. The maximum of the overlap occurs when the impurities are located along the [111] direction, whereas the minimum occurs along the [001] direction.  The TBM and EFM show similar qualitative behavior, however the results differ slightly in value. The EFM zero-range potential model underestimates the magnitude of the wave-function anisotropy compared to that observed in experiment and obtained with TBM.

The wave-function overlap $\Omega{i}^{(xyz)}$ for Mn-Mn pairs grown on $(xyz)$ oriented substrates (Table~\ref{tab:table}) is estimated by averaging the curves~$\Sigma{i}$ for directions perpendicular to $(xyz)$. These calculations suggest that the wave-function overlap on average is different for Mn $\delta$-doped layers grown on differently oriented substrates. The density where the Mott metal-insulator transition occurs in an impurity band is determined by the overlap of localized wave functions. The anisotropic overlap of the Mn wave-functions will produce a directionally-dependent density threshold for percolation as well.   Thus the critical concentration for the metal-insulator transition will be lower for Mn $\delta$-doped layers grown on (111) or (110) substrates compared to $\delta$-doped layers grown on (001) substrates. As the Curie temperatures of metallic \hbox{GaMnAs} are much higher than those of insulating \hbox{GaMnAs}, the Curie temperatures and other magnetic properties for $\delta$-doped layers should be strongly dependent on the substrate orientation, with (111) substrates yielding higher Curie temperatures than (110) substrates or the currently-used (001) 
substrates.

\begin{table}
\caption{\label{tab:table}Calculated values of the Mn wave-function overlap from the EFM. $\Sigma$~is the curve number in figure~\ref{2Mn angle}, \emph{D}~is the Mn-Mn separation, \emph{N}~is the normalization coefficient, and $\Omega^{(xyz)}$~is the averaged overlap integral for Mn pairs grown on a ($xyz$) substrate.}
\begin{ruledtabular}
\begin{tabular}{cccccccc}
 &$D$ (nm) &$N$ &$\Omega^{(001)}$ &$\Omega^{(110)}$ &$\Omega^{(111)}$\\
\hline
$\Sigma0$& 0.5  & 0.654 & 0.986 & 0.990 & 0.992\\
$\Sigma1$& 1 & 0.438 & 0.963 & 0.976 & 0.980\\
$\Sigma2$& 2 & 0.216 & 0.912 & 0.941 & 0.952\\
$\Sigma3$& 3 & 0.112 & 0.870 & 0.913 & 0.929\\
$\Sigma4$& 4 & 0.060 & 0.837 & 0.892 & 0.912\\
$\Sigma5$& 5 & 0.033 & 0.812 & 0.875 & 0.898\\
\end{tabular}
\end{ruledtabular}
\end{table}
In conclusion, we have experimentally demonstrated that the crosslike shape of the Mn persists in groups of Mn with short Mn-Mn separation. This strongly supports the picture of impurity-band conduction and spin-spin coupling at Mn doping densities corresponding to ferromagnetic \hbox{GaMnAs}. We suggest that the anisotropy of the Mn wave-function will substantially influence the carrier density of the Mott metal-insulator transition in Mn $\delta$-doped layers grown on differently oriented substrates. We expect that Mott transition will occur at lower Mn concentrations in layers grown on (111) substrates and at higher concentrations in layers grown on (001) substrates, leading to higher Curie temperatures for (111)-grown  than (001)-grown $\delta$-doped layers. These results have broad implications for all acceptor-acceptor interactions in zincblende semiconductors, and especially for hole-mediated ferromagnetic semiconductors.

This work was supported by the Dutch Foundation for Fundamental Research on Matter
(FOM), the Belgian Fund for Scientific Research Flanders (FWO), the EC GROWTH
project FENIKS (G5RD-CT-2001-00535), and the ARO MURI DAAD-19-01-1-0541.

\bibliography{Mn}

\hyphenation{Post-Script Sprin-ger}
\begin{thebibliography}{28}
\expandafter\ifx\csname natexlab\endcsname\relax\def\natexlab#1{#1}\fi
\expandafter\ifx\csname bibnamefont\endcsname\relax
  \def\bibnamefont#1{#1}\fi
\expandafter\ifx\csname bibfnamefont\endcsname\relax
  \def\bibfnamefont#1{#1}\fi
\expandafter\ifx\csname citenamefont\endcsname\relax
  \def\citenamefont#1{#1}\fi
\expandafter\ifx\csname url\endcsname\relax
  \def\url#1{\texttt{#1}}\fi
\expandafter\ifx\csname urlprefix\endcsname\relax\def\urlprefix{URL }\fi
\providecommand{\bibinfo}[2]{#2}
\providecommand{\eprint}[2][]{\url{#2}}

\bibitem[{\citenamefont{Awschalom et~al.}(2002)\citenamefont{Awschalom,
  Samarth, and Loss}}]{SpintronicsBook}
\bibinfo{editor}{\bibfnamefont{D.~D.} \bibnamefont{Awschalom}},
  \bibinfo{editor}{\bibfnamefont{N.}~\bibnamefont{Samarth}}, \bibnamefont{and}
  \bibinfo{editor}{\bibfnamefont{D.}~\bibnamefont{Loss}}, eds.,
  \emph{\bibinfo{title}{Semiconductor Spintronics and Quantum Computation}}
  (\bibinfo{publisher}{Springer}, \bibinfo{year}{2002}).

\bibitem[{\citenamefont{MacDonald et~al.}(2005)\citenamefont{MacDonald,
  Schiffer, and Samarth}}]{Nitin-review}
\bibinfo{author}{\bibfnamefont{A.~H.} \bibnamefont{MacDonald}},
  \bibinfo{author}{\bibfnamefont{P.}~\bibnamefont{Schiffer}}, \bibnamefont{and}
  \bibinfo{author}{\bibfnamefont{N.}~\bibnamefont{Samarth}},
  \bibinfo{journal}{Nature Materials} \textbf{\bibinfo{volume}{4}},
  \bibinfo{pages}{195} (\bibinfo{year}{2005}).

\bibitem[{\citenamefont{Yakunin
  et~al.}(2004{\natexlab{a}})\citenamefont{Yakunin, {A.~Yu.~Silov}, Koenraad,
  {W.~Van~Roy}, {J.~De~Boeck}, and Wolter}}]{Manipulation}
\bibinfo{author}{\bibfnamefont{A.~M.} \bibnamefont{Yakunin}},
  \bibinfo{author}{\bibnamefont{{A.~Yu.~Silov}}},
  \bibinfo{author}{\bibfnamefont{P.~M.} \bibnamefont{Koenraad}},
  \bibinfo{author}{\bibnamefont{{W.~Van~Roy}}},
  \bibinfo{author}{\bibnamefont{{J.~De~Boeck}}}, \bibnamefont{and}
  \bibinfo{author}{\bibfnamefont{J.~H.} \bibnamefont{Wolter}},
  \bibinfo{journal}{Physica~E} \textbf{\bibinfo{volume}{21}},
  \bibinfo{pages}{947} (\bibinfo{year}{2004}{\natexlab{a}}).

\bibitem[{\citenamefont{Lee and Anderson}(1964)}]{PL_Mn_1964}
\bibinfo{author}{\bibfnamefont{T.~C.} \bibnamefont{Lee}} \bibnamefont{and}
  \bibinfo{author}{\bibfnamefont{W.~W.} \bibnamefont{Anderson}},
  \bibinfo{journal}{Solid State Commun.} \textbf{\bibinfo{volume}{2}},
  \bibinfo{pages}{265} (\bibinfo{year}{1964}).

\bibitem[{\citenamefont{Schneider et~al.}(1987)\citenamefont{Schneider,
  Kaufmann, Wilkening, Baeumler, and K\"{o}hl}}]{Schneider_Mn+h}
\bibinfo{author}{\bibfnamefont{J.}~\bibnamefont{Schneider}},
  \bibinfo{author}{\bibfnamefont{U.}~\bibnamefont{Kaufmann}},
  \bibinfo{author}{\bibfnamefont{W.}~\bibnamefont{Wilkening}},
  \bibinfo{author}{\bibfnamefont{M.}~\bibnamefont{Baeumler}}, \bibnamefont{and}
  \bibinfo{author}{\bibfnamefont{F.}~\bibnamefont{K\"{o}hl}},
  \bibinfo{journal}{Phys. Rev. Lett.} \textbf{\bibinfo{volume}{59}},
  \bibinfo{pages}{240} (\bibinfo{year}{1987}).

\bibitem[{\citenamefont{Averkiev et~al.}(1988)\citenamefont{Averkiev, Gutkin,
  Osipov, and Reshnichov}}]{Averkiev_Mn_J1-4}
\bibinfo{author}{\bibfnamefont{N.~S.} \bibnamefont{Averkiev}},
  \bibinfo{author}{\bibfnamefont{A.~A.} \bibnamefont{Gutkin}},
  \bibinfo{author}{\bibfnamefont{E.~B.} \bibnamefont{Osipov}},
  \bibnamefont{and} \bibinfo{author}{\bibfnamefont{M.~A.}
  \bibnamefont{Reshnichov}}, \bibinfo{journal}{Sov. Phys. Solid State}
  \textbf{\bibinfo{volume}{30}}, \bibinfo{pages}{438} (\bibinfo{year}{1988}),
  \bibinfo{note}{[Fiz. Tverd. Tela, \textbf{30}, 765 (1988)]}.

\bibitem[{\citenamefont{Linnarsson et~al.}(1997)\citenamefont{Linnarsson,
  Janz\'{e}n, Monemar, Kleverman, and Thilderkvist}}]{Linnarsson_Mn_ExSp}
\bibinfo{author}{\bibfnamefont{M.}~\bibnamefont{Linnarsson}},
  \bibinfo{author}{\bibfnamefont{E.}~\bibnamefont{Janz\'{e}n}},
  \bibinfo{author}{\bibfnamefont{B.}~\bibnamefont{Monemar}},
  \bibinfo{author}{\bibfnamefont{M.}~\bibnamefont{Kleverman}},
  \bibnamefont{and}
  \bibinfo{author}{\bibfnamefont{A.}~\bibnamefont{Thilderkvist}},
  \bibinfo{journal}{Phys. Rev. B} \textbf{\bibinfo{volume}{55}},
  \bibinfo{pages}{6938} (\bibinfo{year}{1997}).

\bibitem[{\citenamefont{Yakunin
  et~al.}(2004{\natexlab{b}})\citenamefont{Yakunin, {A.~Yu.~Silov}, Koenraad,
  Wolter, {W.~Van~Roy}, {J.~De~Boeck}, Tang, and Flatt\'{e}}}]{YakuninPRL_Mn}
\bibinfo{author}{\bibfnamefont{A.~M.} \bibnamefont{Yakunin}},
  \bibinfo{author}{\bibnamefont{{A.~Yu.~Silov}}},
  \bibinfo{author}{\bibfnamefont{P.~M.} \bibnamefont{Koenraad}},
  \bibinfo{author}{\bibfnamefont{J.~H.} \bibnamefont{Wolter}},
  \bibinfo{author}{\bibnamefont{{W.~Van~Roy}}},
  \bibinfo{author}{\bibnamefont{{J.~De~Boeck}}},
  \bibinfo{author}{\bibfnamefont{J.~M.} \bibnamefont{Tang}}, \bibnamefont{and}
  \bibinfo{author}{\bibfnamefont{M.~E.} \bibnamefont{Flatt\'{e}}},
  \bibinfo{journal}{Phys. Rev. Lett.} \textbf{\bibinfo{volume}{92}},
  \bibinfo{pages}{216806} (\bibinfo{year}{2004}{\natexlab{b}}).

\bibitem[{\citenamefont{{J.~-M.~Tang} and Flatt\'{e}}(2004)}]{Flatte_Mn}
\bibinfo{author}{\bibnamefont{{J.~-M.~Tang}}} \bibnamefont{and}
  \bibinfo{author}{\bibfnamefont{M.~E.} \bibnamefont{Flatt\'{e}}},
  \bibinfo{journal}{Phys. Rev. Lett.} \textbf{\bibinfo{volume}{92}},
  \bibinfo{pages}{047201} (\bibinfo{year}{2004}).

\bibitem[{\citenamefont{Guo et~al.}(2004)\citenamefont{Guo, Chen, Sun, Zhou,
  Sun, Cao, and Lu}}]{Guo_2004_Mn-Mn}
\bibinfo{author}{\bibfnamefont{X.~G.} \bibnamefont{Guo}},
  \bibinfo{author}{\bibfnamefont{X.~S.} \bibnamefont{Chen}},
  \bibinfo{author}{\bibfnamefont{Y.~L.} \bibnamefont{Sun}},
  \bibinfo{author}{\bibfnamefont{X.~H.} \bibnamefont{Zhou}},
  \bibinfo{author}{\bibfnamefont{L.~Z.} \bibnamefont{Sun}},
  \bibinfo{author}{\bibfnamefont{J.~C.} \bibnamefont{Cao}}, \bibnamefont{and}
  \bibinfo{author}{\bibfnamefont{W.}~\bibnamefont{Lu}}, \bibinfo{journal}{Phys.
  Rev. B} \textbf{\bibinfo{volume}{69}}, \bibinfo{pages}{085206}
  (\bibinfo{year}{2004}).

\bibitem[{\citenamefont{Fiete et~al.}(2003)\citenamefont{Fiete, Zar\'and, and
  Damle}}]{Fiete_Mn-Mn}
\bibinfo{author}{\bibfnamefont{G.~A.} \bibnamefont{Fiete}},
  \bibinfo{author}{\bibfnamefont{G.}~\bibnamefont{Zar\'and}}, \bibnamefont{and}
  \bibinfo{author}{\bibfnamefont{K.}~\bibnamefont{Damle}},
  \bibinfo{journal}{Phys. Rev. Lett.} \textbf{\bibinfo{volume}{91}},
  \bibinfo{pages}{097202} (\bibinfo{year}{2003}).

\bibitem[{\citenamefont{Dietl et~al.}(2000)\citenamefont{Dietl, Ohno,
  Matsukura, Cibert, and Ferrand}}]{Dietl_Mn_Science}
\bibinfo{author}{\bibfnamefont{T.}~\bibnamefont{Dietl}},
  \bibinfo{author}{\bibfnamefont{H.}~\bibnamefont{Ohno}},
  \bibinfo{author}{\bibfnamefont{F.}~\bibnamefont{Matsukura}},
  \bibinfo{author}{\bibfnamefont{J.}~\bibnamefont{Cibert}}, \bibnamefont{and}
  \bibinfo{author}{\bibfnamefont{D.}~\bibnamefont{Ferrand}},
  \bibinfo{journal}{Science} \textbf{\bibinfo{volume}{287}},
  \bibinfo{pages}{1019} (\bibinfo{year}{2000}).

\bibitem[{\citenamefont{Okabayashi et~al.}(2001)\citenamefont{Okabayashi,
  Kimura, Rader, Mizokawa, Fujimori, Hayashi, and
  Tanaka}}]{Okabayashi_AnglRes_PhEm}
\bibinfo{author}{\bibfnamefont{J.}~\bibnamefont{Okabayashi}},
  \bibinfo{author}{\bibfnamefont{A.}~\bibnamefont{Kimura}},
  \bibinfo{author}{\bibfnamefont{O.}~\bibnamefont{Rader}},
  \bibinfo{author}{\bibfnamefont{T.}~\bibnamefont{Mizokawa}},
  \bibinfo{author}{\bibfnamefont{A.}~\bibnamefont{Fujimori}},
  \bibinfo{author}{\bibfnamefont{T.}~\bibnamefont{Hayashi}}, \bibnamefont{and}
  \bibinfo{author}{\bibfnamefont{M.}~\bibnamefont{Tanaka}},
  \bibinfo{journal}{Phys. Rev. B} \textbf{\bibinfo{volume}{64}},
  \bibinfo{pages}{125304} (\bibinfo{year}{2001}).

\bibitem[{\citenamefont{Nagai et~al.}(2001)\citenamefont{Nagai, Kunimoto,
  Nagasaka, Nojri, Motokawa, Matsukura, Dietl, and Ohno}}]{Nagai_SpinPolAdsorb}
\bibinfo{author}{\bibfnamefont{Y.}~\bibnamefont{Nagai}},
  \bibinfo{author}{\bibfnamefont{T.}~\bibnamefont{Kunimoto}},
  \bibinfo{author}{\bibfnamefont{K.}~\bibnamefont{Nagasaka}},
  \bibinfo{author}{\bibfnamefont{H.}~\bibnamefont{Nojri}},
  \bibinfo{author}{\bibfnamefont{M.}~\bibnamefont{Motokawa}},
  \bibinfo{author}{\bibfnamefont{F.}~\bibnamefont{Matsukura}},
  \bibinfo{author}{\bibfnamefont{T.}~\bibnamefont{Dietl}}, \bibnamefont{and}
  \bibinfo{author}{\bibfnamefont{H.}~\bibnamefont{Ohno}},
  \bibinfo{journal}{Jpn. J. Appl. Phys.} \textbf{\bibinfo{volume}{40}},
  \bibinfo{pages}{6231} (\bibinfo{year}{2001}).

\bibitem[{\citenamefont{Singley et~al.}(2002)\citenamefont{Singley, Kawakami,
  Awschalom, and Basov}}]{Singley_Infrared}
\bibinfo{author}{\bibfnamefont{E.~J.} \bibnamefont{Singley}},
  \bibinfo{author}{\bibfnamefont{R.}~\bibnamefont{Kawakami}},
  \bibinfo{author}{\bibfnamefont{D.~D.} \bibnamefont{Awschalom}},
  \bibnamefont{and} \bibinfo{author}{\bibfnamefont{D.~N.} \bibnamefont{Basov}},
  \bibinfo{journal}{Phys. Rev. Lett.} \textbf{\bibinfo{volume}{89}},
  \bibinfo{pages}{097203} (\bibinfo{year}{2002}).

\bibitem[{\citenamefont{Sapega et~al.}(2002)\citenamefont{Sapega, Moreno,
  Ramsteiner, D\"{a}weritz, and Ploog}}]{SapegaMn}
\bibinfo{author}{\bibfnamefont{V.~F.} \bibnamefont{Sapega}},
  \bibinfo{author}{\bibfnamefont{M.}~\bibnamefont{Moreno}},
  \bibinfo{author}{\bibfnamefont{M.}~\bibnamefont{Ramsteiner}},
  \bibinfo{author}{\bibfnamefont{L.}~\bibnamefont{D\"{a}weritz}},
  \bibnamefont{and} \bibinfo{author}{\bibfnamefont{K.}~\bibnamefont{Ploog}},
  \bibinfo{journal}{Phys. Rev. B} \textbf{\bibinfo{volume}{66}},
  \bibinfo{pages}{075217} (\bibinfo{year}{2002}).

\bibitem[{\citenamefont{Berciu and Bhatt}(2001)}]{Bhatt-PRL-2001}
\bibinfo{author}{\bibfnamefont{M.}~\bibnamefont{Berciu}} \bibnamefont{and}
  \bibinfo{author}{\bibfnamefont{R.~N.} \bibnamefont{Bhatt}},
  \bibinfo{journal}{Phys. Rev. Lett.} \textbf{\bibinfo{volume}{87}},
  \bibinfo{pages}{107203} (\bibinfo{year}{2001}).

\bibitem[{\citenamefont{Timm et~al.}(2002)\citenamefont{Timm, Sch\"afer, and
  von Oppen}}]{Timm-PRL-2002}
\bibinfo{author}{\bibfnamefont{C.}~\bibnamefont{Timm}},
  \bibinfo{author}{\bibfnamefont{F.}~\bibnamefont{Sch\"afer}},
  \bibnamefont{and} \bibinfo{author}{\bibfnamefont{F.}~\bibnamefont{von
  Oppen}}, \bibinfo{journal}{Phys. Rev. Lett.} \textbf{\bibinfo{volume}{89}},
  \bibinfo{pages}{137201} (\bibinfo{year}{2002}).

\bibitem[{\citenamefont{Fiete et~al.}(2005)\citenamefont{Fiete, Zar\'and,
  Damle, and Moca}}]{Fiete_GaMnAs-cm}
\bibinfo{author}{\bibfnamefont{G.~A.} \bibnamefont{Fiete}},
  \bibinfo{author}{\bibfnamefont{G.}~\bibnamefont{Zar\'and}},
  \bibinfo{author}{\bibfnamefont{K.}~\bibnamefont{Damle}}, \bibnamefont{and}
  \bibinfo{author}{\bibfnamefont{C.~P.} \bibnamefont{Moca}},
  \bibinfo{journal}{cond-mat/0503382}  (\bibinfo{year}{2005}).

\bibitem[{\citenamefont{Bouzerar et~al.}(2005)\citenamefont{Bouzerar, Ziman,
  and Kudrnovsk\'y}}]{Bouzerar_Mn}
\bibinfo{author}{\bibfnamefont{G.}~\bibnamefont{Bouzerar}},
  \bibinfo{author}{\bibfnamefont{T.}~\bibnamefont{Ziman}}, \bibnamefont{and}
  \bibinfo{author}{\bibfnamefont{J.}~\bibnamefont{Kudrnovsk\'y}},
  \bibinfo{journal}{Europhys. Lett.} \textbf{\bibinfo{volume}{69}},
  \bibinfo{pages}{812} (\bibinfo{year}{2005}).

\bibitem[{\citenamefont{Tsuruoka
  et~al.}(2002{\natexlab{a}})\citenamefont{Tsuruoka, Tachikawa, Ushioda,
  Matsukura, Takamura, and Ohno}}]{Ohno-XSTM-APL}
\bibinfo{author}{\bibfnamefont{T.}~\bibnamefont{Tsuruoka}},
  \bibinfo{author}{\bibfnamefont{N.}~\bibnamefont{Tachikawa}},
  \bibinfo{author}{\bibfnamefont{S.}~\bibnamefont{Ushioda}},
  \bibinfo{author}{\bibfnamefont{F.}~\bibnamefont{Matsukura}},
  \bibinfo{author}{\bibfnamefont{K.}~\bibnamefont{Takamura}}, \bibnamefont{and}
  \bibinfo{author}{\bibfnamefont{H.}~\bibnamefont{Ohno}},
  \bibinfo{journal}{Appl. Phys. Lett.} \textbf{\bibinfo{volume}{81}},
  \bibinfo{pages}{2800} (\bibinfo{year}{2002}{\natexlab{a}}).

\bibitem[{\citenamefont{Sullivan et~al.}(2003)\citenamefont{Sullivan, Boishin,
  Whitman, Hanbicki, Jonker, and Erwin}}]{NRL-XSTM-PRB}
\bibinfo{author}{\bibfnamefont{J.~M.} \bibnamefont{Sullivan}},
  \bibinfo{author}{\bibfnamefont{G.~I.} \bibnamefont{Boishin}},
  \bibinfo{author}{\bibfnamefont{L.~J.} \bibnamefont{Whitman}},
  \bibinfo{author}{\bibfnamefont{A.~T.} \bibnamefont{Hanbicki}},
  \bibinfo{author}{\bibfnamefont{B.~T.} \bibnamefont{Jonker}},
  \bibnamefont{and} \bibinfo{author}{\bibfnamefont{S.~C.} \bibnamefont{Erwin}},
  \bibinfo{journal}{Phys. Rev. B} \textbf{\bibinfo{volume}{68}},
  \bibinfo{pages}{235324} (\bibinfo{year}{2003}).

\bibitem[{\citenamefont{Grandidier et~al.}(2000)\citenamefont{Grandidier, Nys,
  Delerue, Sti\'evenard, Higo, and Tanaka}}]{Grandidier-XSTM-APL}
\bibinfo{author}{\bibfnamefont{B.}~\bibnamefont{Grandidier}},
  \bibinfo{author}{\bibfnamefont{J.~P.} \bibnamefont{Nys}},
  \bibinfo{author}{\bibfnamefont{C.}~\bibnamefont{Delerue}},
  \bibinfo{author}{\bibfnamefont{D.}~\bibnamefont{Sti\'evenard}},
  \bibinfo{author}{\bibfnamefont{Y.}~\bibnamefont{Higo}}, \bibnamefont{and}
  \bibinfo{author}{\bibfnamefont{M.}~\bibnamefont{Tanaka}},
  \bibinfo{journal}{Appl. Phys. Lett.} \textbf{\bibinfo{volume}{77}},
  \bibinfo{pages}{4001} (\bibinfo{year}{2000}).

\bibitem[{\citenamefont{Tsuruoka
  et~al.}(2002{\natexlab{b}})\citenamefont{Tsuruoka, Tanimoto, Tachikawa,
  Ushioda, Matsukura, and Ohno}}]{Ohno-XSTM}
\bibinfo{author}{\bibfnamefont{T.}~\bibnamefont{Tsuruoka}},
  \bibinfo{author}{\bibfnamefont{R.}~\bibnamefont{Tanimoto}},
  \bibinfo{author}{\bibfnamefont{N.}~\bibnamefont{Tachikawa}},
  \bibinfo{author}{\bibfnamefont{S.}~\bibnamefont{Ushioda}},
  \bibinfo{author}{\bibfnamefont{F.}~\bibnamefont{Matsukura}},
  \bibnamefont{and} \bibinfo{author}{\bibfnamefont{H.}~\bibnamefont{Ohno}},
  \bibinfo{journal}{Solid State Comm.} \textbf{\bibinfo{volume}{121}},
  \bibinfo{pages}{79} (\bibinfo{year}{2002}{\natexlab{b}}).

\bibitem[{\citenamefont{Vogl and Baranowski}(1985)}]{Vogl}
\bibinfo{author}{\bibfnamefont{P.}~\bibnamefont{Vogl}} \bibnamefont{and}
  \bibinfo{author}{\bibfnamefont{J.~M.} \bibnamefont{Baranowski}},
  \bibinfo{journal}{Acta Phys. Pol. A} \textbf{\bibinfo{volume}{67}},
  \bibinfo{pages}{133} (\bibinfo{year}{1985}).

\bibitem[{\citenamefont{Flatt\'e and Reynolds}(2000)}]{Flatte-Reynolds}
\bibinfo{author}{\bibfnamefont{M.~E.} \bibnamefont{Flatt\'e}} \bibnamefont{and}
  \bibinfo{author}{\bibfnamefont{D.~E.} \bibnamefont{Reynolds}},
  \bibinfo{journal}{Phys. Rev. B} \textbf{\bibinfo{volume}{61}},
  \bibinfo{pages}{14810} (\bibinfo{year}{2000}).

\bibitem[{\citenamefont{Averkiev and {S.~Yu.~Il'inskii}}(1994)}]{ZeroRange}
\bibinfo{author}{\bibfnamefont{N.~S.} \bibnamefont{Averkiev}} \bibnamefont{and}
  \bibinfo{author}{\bibnamefont{{S.~Yu.~Il'inskii}}}, \bibinfo{journal}{Phys.
  Solid. State} \textbf{\bibinfo{volume}{36}}, \bibinfo{pages}{278}
  (\bibinfo{year}{1994}), \bibinfo{note}{[Fiz. Tverd. Tela, \textbf{36}, 503
  (1994)]}.

\bibitem[{\citenamefont{Bir and Pikus}(1974)}]{BirPikus}
\bibinfo{author}{\bibfnamefont{G.~L.} \bibnamefont{Bir}} \bibnamefont{and}
  \bibinfo{author}{\bibfnamefont{G.~E.} \bibnamefont{Pikus}},
  \emph{\bibinfo{title}{Symmetry and strain-induced effects in semiconductors}}
  (\bibinfo{publisher}{Halsted}, \bibinfo{address}{Jerusalem},
  \bibinfo{year}{1974}).

\end{thebibliography}


\end{document}